\documentclass[twocolumn,showpacs,preprintnumbers,amsmath,amssymb]{revtex4}
\usepackage{yfonts}
\usepackage{bm}% bold math
\usepackage{amssymb}
\usepackage{graphicx}
\usepackage{color}
\usepackage{wrapfig,subfigure}
\swabfamily

\def\tr{t_{\rm r}}

\newcommand{\pb}{{\bf p}}
\newcommand{\xb}{{\bf x}}
\newcommand{\rb}{{\bf r}}
\newcommand{\Xb}{{\bf X}}
\newcommand{\xbS}{{\bf x_{\cal S}}}
\newcommand{\XbS}{{\bf X_{\cal S}}}
\newcommand{\pbS}{{\bf p_{\cal S}}}

\newcommand{\ka}{\varkappa}

\def\p12{p_{12}({\bf q},t)}

\begin{document}
\title{
Extinction rate fragility in population dynamics
}
\author{M. Khasin and M. I. Dykman}
\affiliation{Department of Physics and Astronomy,
Michigan State University, East Lansing, MI 48824}

\date{\today }

\begin{abstract}
Population extinction is a rare event which requires overcoming an effective barrier.  We show that the extinction rate can be fragile: a small change in the system parameters leads to an exponentially strong change of the rate, with the barrier height depending on the parameters nonanalytically. General conditions of the fragility are established. The fragility is found in one of the best-known models of epidemiology, the  SIS model. The analytical expressions are compared with simulations.
\end{abstract}

\maketitle

Extinction of a population is of central interest for population dynamics \cite{Bartlett1960,Andersson2000}. It results from a large fluctuation away from a steady state of the population. Such fluctuations are usually rare. They require an unlikely sequence of elementary birth-death events or a large change in the fluctuating environment, or both, often visualized as overcoming a barrier. Much work has been done on extinction for various fluctuation mechanisms, and the extinction rates have been found for a number of models of population dynamics \cite{Weiss1971,Leigh1981,Roozen1989,Jacquez1993,Herwaarden1995,Allen2000,Elgart2004,Doering2005,Kessler2007,Dykman2008,Kamenev2008b}.

There is a close similarity between population extinction and a diverse group of physical phenomena which involve switching between coexisting states and range from switching in Josephson junctions and nanomagnets to chemical reactions and to protein folding. Both extinction and switching are caused by large rare fluctuations. In many cases the rate of extinction (switching) $W$ is exponentially small, $W\propto \exp(-{\cal Q})$ with ${\cal Q}\gg 1$  \cite{Freidlin_book}. In particular, for systems close to thermal equilibrium the switching exponent ${\cal Q}$ is ${\cal Q}=R/k_BT$, where $R$ is the free energy barrier and $T$ is temperature \cite{Kramers1940}.

In this paper we show that, in spite of the aforementioned similarity, population extinction displays a feature that does not generally occur in switching between metastable states. We find that the extinction rate is often {\em fragile}. A small perturbation of the system can lead to an abrupt change of the rate exponent ${\cal Q}$. If the perturbation is proportional to a parameter $\mu$, the value of ${\cal Q}$ for $\mu=0$ is much larger than for $\mu\to 0$. We find a general condition for the fragility to occur and illustrate the effect with a broadly used epidemiological model, the so called SIS model where there are present only  susceptible ($S$) and infected ($I$) individuals \cite{Andersson2000}.

\begin{figure}[h]
\includegraphics[width=3.2in]{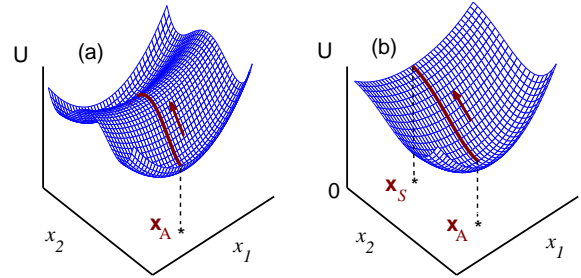}
\caption{A sketch of switching between stable states (a) and extinction (b); $\xb$ are the dynamical variables of the system. The stable states correspond to the minima of the effective potential $U(\xb)$. In switching, a quasi-stationary probability current shown by arrows goes from the initially occupied to the initially empty state over the saddle point. In extinction, the probability current terminates at the extinction state and the occupation of this state linearly increases in time for $t\ll W^{-1}$, where $W$ is the extinction rate.}
\label{fig:sketch_switch}
\end{figure}

The difference between switching and extinction can be understood from Fig.~\ref{fig:sketch_switch}. For illustration purpose, systems that can switch or display extinction are sketched as particles with dynamical variables $\xb$ moving in a potential $U(\xb)$, even though the actual system motion is generally non-potential. In population dynamics, the components of $\xb$ determine the size of different populations.  In switching, if the system is initially near a stable state $\xb_A$, over the relaxation time $\tr$ there is formed a  quasi-stationary current away from the basin of attraction to $\xb_A$. The current gives the switching rate \cite{Kramers1940}. It goes over the saddle point \cite{Kramers1940,Freidlin_book,Day1987a,Dykman1990,Maier1997} and is divergence-free there, the probability distribution does not accumulate near the saddle point.

\begin{figure}[h]
\includegraphics[width=3.2in]{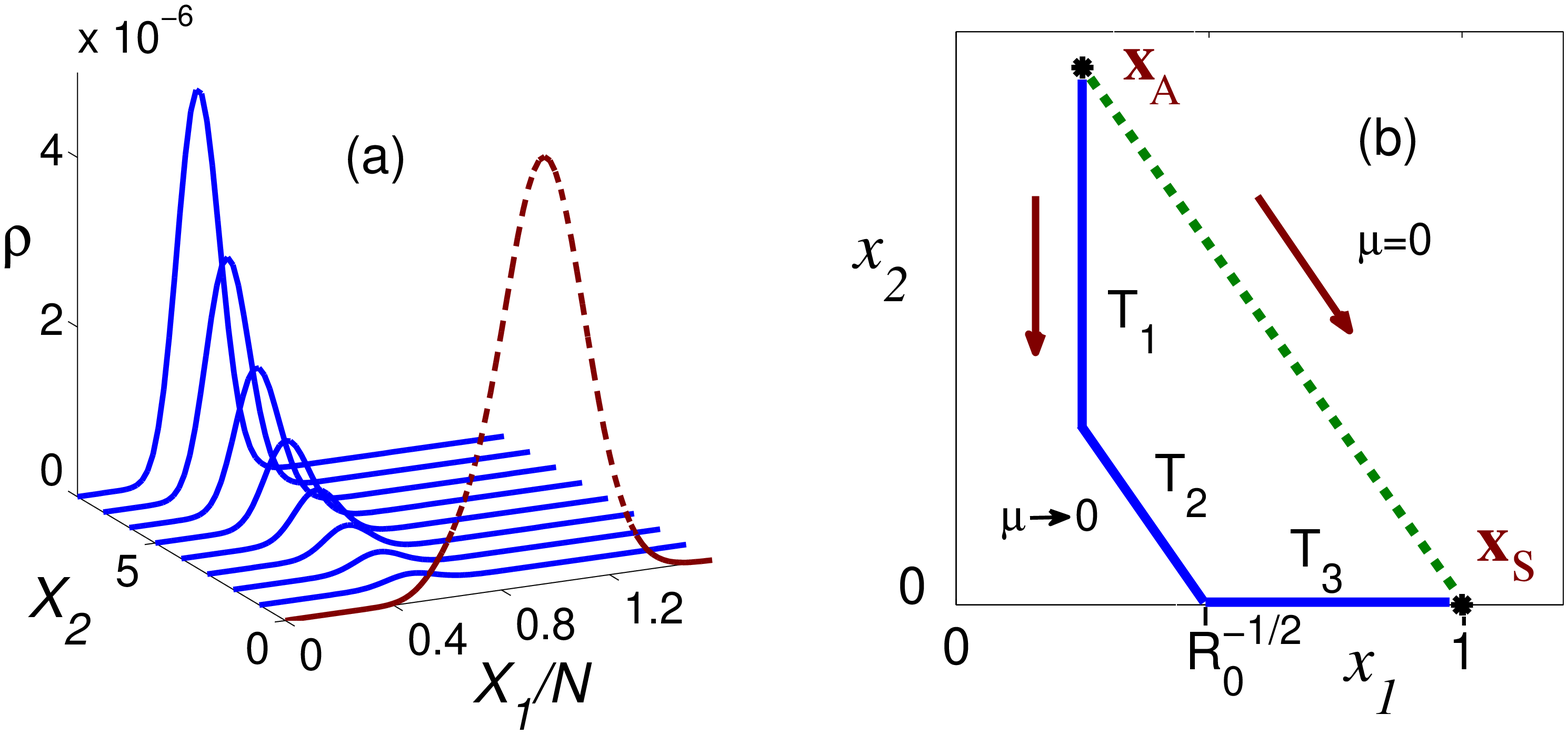}
\caption{ (Color online). (a) A snapshot of the probability $\rho(\Xb)$ near the extinction plane $X_2=0$ for the SIS model; $\rho$ is quasi-continuous in $X_1/N$. The data of simulations refer to $\mu t = 9, R_0=4,\mu'\equiv\mu/(\mu+\kappa)=0.1$. For $t=0$ the system was at $\Xb_A$, the total number of particles was $N=50$. (b) Asymptotic optimal Hamiltonian trajectories for extinction for $\mu\to0$ (solid line) and $\mu=0$ (dashed line). }
\label{fig:optimal_paths}
\end{figure}

The extinction rate is also given by the quasi-stationary current away from the vicinity of the stable state, see Fig.~\ref{fig:sketch_switch}~(b). The current goes to the hyperplane $x_E=0$ where population $E$ goes extinct and, in a qualitative difference from the switching current, {\em terminates} there, because the population size may not become negative. Since the extinct population usually does not emerge again, fluctuations do not remove the system from the hyperplane $x_E=0$. Therefore the probability distribution accumulates there, as seen in Fig.~\ref{fig:optimal_paths}(a) below. For $x_E>0$ the probability distribution is quasi-stationary on times $\tr \ll t \ll W^{-1}$.

Because of the current discontinuity, near $x_E=0$ the quasi-stationary distribution sharply varies with $x_E$ (exponentially sharply, see below). However, with respect to non-extinct populations, $x_{i\neq E}$, it generally has a smooth maximum, cf. Fig.~\ref{fig:optimal_paths}(a). This is no longer true if the system has effective constraints, for example the total population is conserved. Population conservation is a commonly used assumption in the SIS model \cite{Andersson2000,Weiss1971,Leigh1981,Doering2005,Kamenev2008b}, and indeed the distribution over $x_{i\neq E}$ is sharp near $x_E=0$ in the SIS model with constant population. Once the constraint is lifted, for example, the population starts fluctuating, the shape of the distribution changes as does also the extinction rate. The rate change occurs in an exponentially narrow parameter range and is exponentially large, which is a signature of the fragility.

We consider extinction in a spatially uniform system of coupled populations (species). The system state is characterized by a vector $\Xb$ with integer components equal to the size of different populations. The quasi-continuous vector $\xb=\Xb/N$ introduced above gives the populations size scaled by a large characteristic total population $N$. The dynamics is quite generally described by a master equation for the probability $\rho(\Xb)$,
\begin{eqnarray}
\label{eq:master_master}
\dot \rho({\bf X})=\sum_r\left[W({\bf X-r; r})\rho({\bf X-r}) - W({\bf X; r})\rho({\bf X})\right].
\end{eqnarray}
Here, $W(\Xb;\rb)$ is the rate of an elementary transition $\Xb \to \Xb + \rb$ in which the populations change by $\rb=(r_1,r_2,\ldots)$. The condition that the system does not leave the extinction hyperplane has a form
\begin{equation}
\label{eq:ext_plane_condition}
W(\Xb;\rb)=0 \qquad {\rm for} \qquad X_E=0,\; r_E\neq 0.
\end{equation}

If fluctuations can be disregarded, from Eq.~(\ref{eq:master_master}) we obtain for average scaled populations $\bar\xb$ a mean-field equation
\begin{equation}
\label{eq:mean_field}
\dot{\bar\xb} =\sum\nolimits_{\rb}\rb w(\bar\xb;\rb),
\end{equation}
where  $w(\xb;\rb)=W(\Xb;\rb)/N$ is a characteristic transition rate per individual. We assume that Eq.~(\ref{eq:mean_field}) has an asymptotically stable solution $\xb_A$ and a stationary solution $\xbS$ that lies on the extinction hyperplane $x_E=0$, cf. Fig.~\ref{fig:sketch_switch}(b), and is asymptotically stable with respect to $x_{i\neq E}$ but unstable for $x_E$. For $\tr\ll t\ll W^{-1}$ the distribution $\rho(\Xb)$ peaks at $\Xb_A=N\xb_A$.

The exponent ${\cal Q}$ in the extinction rate $W\propto \exp{\cal Q}$ can be found by either solving the mean first passage time problem for reaching extinction \cite{Weiss1971,Leigh1981,Herwaarden1995} or by calculating the small-$X_E$ tail of $\rho(\Xb)$ \cite{Elgart2004,Doering2005,Dykman2008,Kamenev2008b}. In both methods one looks for the optimal (most probable) fluctuation that leads to extinction, and the results coincide. Here we will study the quasi-stationary distribution. In a standard way, we seek the solution of Eq.~(\ref{eq:master_master}) in the eikonal form,
\begin{eqnarray}
\label{eq:eikonal_approximation}
&&\rho({\bf X})=\exp[-Ns({\bf x})], \qquad \dot s=-H(\xb,\partial_{\xb}s), \nonumber\\
&& H(\xb,\pb)=\sum\nolimits_{\bf r}w(\xb;\rb)\left[\exp(\pb\rb)-1\right].
\end{eqnarray}
We took into account that, typically, $|\rb|\ll N$ and $W(\Xb;\rb)$ depends on $\Xb$ polynomially, whereas $\rho$ is exponential in $\Xb$. Therefore we expanded $\rho({\bf X+r})\approx \rho({\bf X})\exp(-\rb\partial_{\xb}s)$ and replaced $w(\xb-\rb/N;\rb)\to w(\xb;\rb)$.

Equation (\ref{eq:eikonal_approximation}) reduces the problem of the quasi-stationary probability distribution to the problem of classical dynamics of an auxiliary Hamiltonian system with equations of motion
\begin{eqnarray}
\label{eq:eom_Hamiltonian}
\dot\xb=\sum\nolimits_{\rb}\rb w(\xb;\rb)e^{\pb\rb}, \quad \dot \pb=-\sum\nolimits_{\rb}\partial_{\xb} w(\xb;\rb)\left(e^{\pb\rb}-1\right).
\end{eqnarray}
The distribution $\rho(\Xb)$ is determined by the mechanical action of the auxiliary system $s(\xb)$. In the quasi-stationary regime $\dot s=0$, i.e., $H=0$ in Eq.~(\ref{eq:eikonal_approximation}). We notice that $\sum\nolimits_{\rb} w(\xb;\rb)\rb\partial_{\xb}s \leq 0$ for $H(\xb,\partial_{\xb}s)=0$, and therefore $s(\xb)$ decreases if the point $\xb$ shifts along a mean-field trajectory (\ref{eq:mean_field}) [provided $\partial_{\xb}s \neq {\bf 0}$]  \cite{Gang1987,Dykman1994d}. Since the mean-field trajectories go to $\xb_A$, the action $s(\xb)$ is minimal at $\xb_A$. Respectively, $\rho(\Xb)$ is maximal for $\Xb_A=N\xb_A$, as expected on physical grounds.

In the spirit of the method of optimal fluctuation \cite{Freidlin_book,Kamenev2008b}, the extinction rate exponent is determined by the minimum of $s(\xb)$ on the extinction hyperplane. From the above arguments, the minimum is reached at the extinction state $\xbS$. Therefore
\begin{equation}
\label{eq:general_Q}
{\cal Q}=N\left[s(\xbS)-s(\xb_A)\right]=N\int\nolimits_{-\infty}^{\infty}dt \,\pb\dot\xb.
\end{equation}
Equation~(\ref{eq:general_Q}) corresponds to the intuitive picture in which the most probable fluctuation leading to extinction starts from the stable state and brings the system to the extinction state, cf. Fig.~\ref{fig:sketch_switch}. The respective optimal Hamiltonian trajectory, Eq.~(\ref{eq:eom_Hamiltonian}), goes from the Hamiltonian fixed point $(\xb_A,\pb= {\bf 0})$ to the fixed point $(\xbS,\pbS)$.

We now find the final momentum $\pbS$. Since action $s(\xb)$ is maximal with respect to $x_{i\neq E}$ at $\xbS$,  $(p_{\cal S})_{i\neq E} = 0$. To find $(p_{\cal S})_E$ we note that, if $w(\xb;\rb)$ smoothly vary with $\xb$, then quite generally, from Eq.~(\ref{eq:ext_plane_condition}) $w(\xb;\rb)\propto x_E$ for $r_E\neq 0$, and from $H=0$
\begin{equation}
\label{eq:E_momdentum_component}
\sum_{\rb, \,r_E\neq 0}\left[x_E^{-1}w(\xb;\rb)\right]_{\xb\to\xbS}\left\{\exp\left[(p_{\cal S})_Er_E\right]-1\right\} =0.
\end{equation}

Equation~(\ref{eq:E_momdentum_component}) has a trivial solution $(p_{\cal S})_E=0$. However, there are no Hamiltonian trajectories that would go from $(\xb_A,\pb={\bf 0})$ to $(\xbS, \pb={\bf 0})$. Indeed, using Eq.~(\ref{eq:mean_field}) one can show that trajectories that go to $(\xbS, \pb={\bf 0})$ lie on the manifold $x_E=0, p_{i\neq E}=0$. This manifold does not contain the point $(\xb_A,\pb={\bf 0})$. The trajectory that gives the exponent ${\cal Q}$ goes to $(\xbS,\pbS)$ with  $(p_{\cal S})_E\neq 0$, as found earlier for specific models \cite{Herwaarden1995,Elgart2004,Dykman2008}. Therefore near $\xbS$ the quasi-stationary distribution $\rho$ as a function of $\xb$ steeply varies with $x_E$, $\rho\propto\exp[-N(p_{\cal S})_Ex_E]$.

The above analysis can be extended to systems with effective constraints, which are implicit in functions $w(\xb;\rb)$ and, for example, give extra integrals of motion. In this case the above conclusions about $\pbS$ change; in particular it is no longer necessary to have $(p_{\cal S})_{i\neq E}=0$. To gain intuition into this change and its dramatic effect on ${\cal Q}$ we will consider the problem of disease extinction in the SIS model. In this model the numbers of susceptible and infected individuals $X_1$ and $X_2$ change because of birth and death, with rates
\begin{eqnarray}
\label{eq:birth-death}
&&W\bigl({\bf X}; (1, 0)\bigr)=N\mu,\quad W\bigl({\bf X}; (-1, 0)\bigr)=\mu X_1,\nonumber\\
&&W\bigl({\bf X}; (0, -1)\bigr)=\mu X_2,
\end{eqnarray}
and because of infection on contact and recovery, with those recovered immediately becoming susceptible \cite{Andersson2000}. The corresponding rates are
\begin{eqnarray}
\label{eq:infection_rates}
W\bigl({\bf X}; (-1, 1)\bigr)=\beta X_1X_2/N, \quad W\bigl({\bf X}; (1, -1)\bigr)=\varkappa X_2.
\end{eqnarray}
Disease extinction occurs where $X_2\equiv X_E=0$. For the infection reproductive rate $R_0>1$, where $R_0=\beta/(\mu+\varkappa)$, the system has an endemic equilibrium $\xb_A=\Xb_A/N=(R_0^{-1}, 1-R_0^{-1})$ . It coexists with the disease-free stationary state $\xbS=\XbS/N=(1,0)$.

Much work has been done on the SIS model in the limit $\mu=0$ where the total population does not fluctuate, $x_1+x_2=1$ \cite{Andersson2000,Weiss1971,Leigh1981,Jacquez1993,Doering2005}. Here, the Hamiltonian system (\ref{eq:eom_Hamiltonian}) has effectively one degree of freedom. A direct substitution shows that on the optimal trajectory $p_2=0$ and $p_1=\ln(\beta x_1/\varkappa)$, which gives
\begin{equation}
\label{eq:mu=0}
{\cal Q}_{\mu=0}=N\left(\ln R_0-1+R_0^{-1}\right).
\end{equation}
In this case $(p_{\cal S})_E\equiv (p_{\cal S})_2=0$ whereas $(p_{\cal S})_1=\ln R_0$. This is in contradiction with the general result for extinction in unconstrained systems and is a consequence of the conservation of the total population.

We now consider the situation where the total population is fluctuating, albeit slowly, that is the characteristic birth-death rate $\mu\ll \varkappa$. Still we assume that $\mu\gg W$, so that the distribution is quasi-stationary. The Hamiltonian trajectory for extinction consists of three almost straight sections $T1, T2, T3$ shown in Fig.~\ref{fig:optimal_paths}(b). Sections $T1, T3$ correspond to slow motion characterized by time $\mu^{-1}$, whereas motion in section $T2$ is fast, with typical time $(\beta-\ka)^{-1}$.  A direct substitution shows that motion in section $T1$ is described by equations
\begin{eqnarray}
\label{eq:eom_T1}
p_1= p_2= \ln\left[1+e^{\mu(t-t_1)}\right],\qquad x_2 = e^{p_2}-R_0^{-1},
\end{eqnarray}
while $|x_1-R_0^{-1}|\lesssim \mu$ [$t_1$ in Eq.~(\ref{eq:eom_T1}) is arbitrary].

Motion in section $T2$ can be described by setting $\mu=0$ in Eqs.~(\ref{eq:eom_Hamiltonian}), (\ref{eq:birth-death}), (\ref{eq:infection_rates}). This gives
\begin{eqnarray}
\label{eq:eom_T2}
p_2=\ln C, \qquad p_1=\ln(CR_0x_1), \qquad x_1+x_2=C,
\end{eqnarray}
where $x_1=(Cf_1+\ka)/(\beta +f_1)$ with $f_1=\exp[(\beta C-\ka)(t-t_2)]$; constants $C,t_2$ should be found by matching the solutions given by Eqs.~(\ref{eq:eom_T1}), (\ref{eq:eom_T2}); . If we set $C=R_0^{-1/2}$, then $x_1\to R_0^{-1/2}, x_2\to 0$, and $p_1\to 0$ for $t\to\infty$, and the trajectory approaches section $T3$. On section $T3$
\begin{eqnarray}
\label{eq:eom_T3}
p_2= -\ln(R_0x_1),\qquad  x_1= 1-\exp[-\mu(t-t_3)],
\end{eqnarray}
while $|p_1|,x_2\to 0$ for $\mu\to 0$.

The solutions match if at the end of section $T1$ and at the beginning of section $T2$ we have $p_1=p_2=-(\ln R_0)/2$. At the end of section $T3$ we have $\xb\to \xbS=(1,0)$ and $\pb\to \pbS=(0,-\ln R_0)$, as expected from the general analysis of unconstrained extinction problem. The extinction rate exponent is
\begin{equation}
\label{eq:Q_mu_to_0}
{\cal Q}_{\mu \to 0}=N(R_0^{1/2}-1)^2/R_0.
\end{equation}
This value, which is obtained in the limit $\mu \to 0$, is smaller than ${\cal Q}$  for $\mu=0$, cf. Eq.~(\ref{eq:mu=0}). The discontinuity with respect to $\mu$ shows the fragility of the result obtained by disregarding fluctuations of the total population.

\begin{figure}[h]
\includegraphics[width=2.4in]{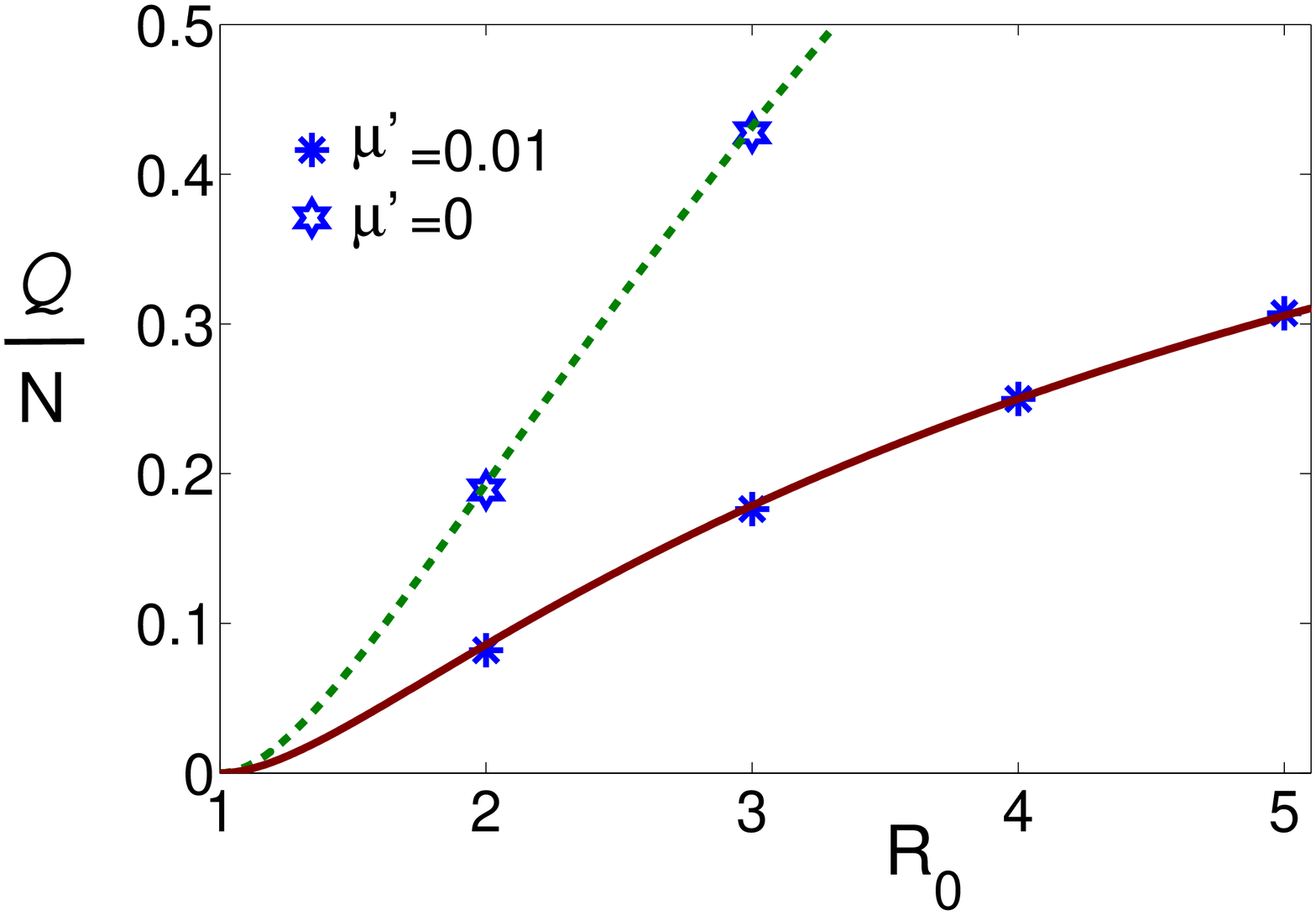}
\caption{ (Color online). The switching exponent ${\cal Q}$ for the SIS model of epidemics. The solid and dashed lines show the results for $\mu\to 0$ [Eq.~(\ref{eq:Q_mu_to_0})] and $\mu=0$ [Eq.~(\ref{eq:mu=0})], respectively. The data points are obtained from the numerical solution of the master equation for the total initial populations $N=50$ and $N=100$, which made it possible to directly extract the exponent ${\cal Q}$. }
\label{fig:comparison}
\end{figure}

In Fig.~\ref{fig:comparison} we compare the values of ${\cal Q}$ obtained for $\mu=0$ and for $\mu\to 0$. Also shown are the results of numerical simulations obtained for $\mu=0$ and for small nonzero $\mu$. They are in excellent agreement with the analytical results. As illustrated in Fig.~\ref{fig:optimal_paths}(a), for $Wt\ll 1$ the probability distribution accumulates linearly in time near $x_1=X_1/N=1$ in the extinction plane, $X_2=0$. Away from the extinction plane, for discrete $X_2\geq 1$, the distribution is quasi-stationary. For small $X_2$ it has a peak along $x_1$ at $\approx R_0^{-1/2}$ where the asymptotic extinction path approaches the plane $x_2=0$.

The fragility in the SIS model results from the inapplicability of a perturbation theory in the fluctuations of the total population. We now show that a perturbation theory quite generally breaks down in the problem of extinction in systems with effective constraints (integrals of motion); the fragility in the SIS model follows from this analysis. We assume that, because of the constraint, on the optimal extinction trajectory $(p_{\cal S})_{i\neq E}\neq 0$ at least for one $i$. A perturbation changes the elementary transition rates in Eq.~(\ref{eq:master_master}), $W(\Xb;\rb)\to W(\Xb;\rb)+\mu W^{(1)}(\Xb;\rb)$, with $\mu\ll 1$ for a small perturbation. The Hamiltonian in the eikonal approximation for the probability distribution, Eq.~(\ref{eq:eikonal_approximation}), is respectively modified, $H\to H+\mu H^{(1)}$. To first order in $\mu$ the resulting change of the extinction exponent ${\cal Q}^{(1)}$ can be calculated along the trajectory $\xb(t),\pb(t)$ of the unperturbed Hamiltonian \cite{LL_Mechanics2004},
\begin{eqnarray}
\label{eq:perturbation_general}
&&{\cal Q}^{(1)}=-N\mu\int\nolimits_{-\infty}^{\infty}dt H^{(1)}\bigl(\xb(t),\pb(t)\bigr)\\
&&H^{(1)}(\xb,\pb)=\sum\nolimits_{\bf r}w^{(1)}(\xb;\rb)\left[\exp(\pb\rb)-1\right].\nonumber
\end{eqnarray}
Because $\pb(t)$ exponentially decays for $t\to -\infty$ (where $\xb\to \xb_A$), the integral over time in Eq.~(\ref{eq:perturbation_general}) does not diverge at the lower limit.

In order for the perturbation not to destroy the extinction state, $W^{(1)}$ must satisfy condition (\ref{eq:ext_plane_condition}), and then quite generally $w^{(1)}=W^{(1)}/N\propto x_E$ for $x_E\to 0$ and $r_E\neq 0$. Since on the optimal extinction trajectory $x_E(t) $ exponentially decays for $t\to \infty$, the integral (\ref{eq:perturbation_general}) of the terms with $r_E\neq 0$ in $H^{(1)}$ converges on the upper limit. However, because by assumption at the endpoint of the optimal trajectory $p_{i\neq E}\neq 0$  for some $i$, and because generally
\begin{equation}
\label{eq:fragility_condition}
w^{(1)}(\xbS;\rb) \not\equiv 0 \qquad {\rm for} \qquad r_E=0,
\end{equation}
the Hamiltonian $H^{(1)}$ remains nonzero for $t\to \infty$ and overall the integral Eq.~(\ref{eq:perturbation_general}) diverges. 

The divergence of ${\cal Q}^{(1)}$ means that the optimal extinction trajectory changes nonperturbatively, as does also the rate exponent ${\cal Q}$, i.e., the extinction rate is fragile with respect to the corresponding perturbation. Population fluctuations in the SIS model provide an example of such a perturbation, as seen from the comparison of Eqs.~(\ref{eq:birth-death}) and (\ref{eq:fragility_condition}). We note that the divergence of the perturbation theory does not emerge in the problem of switching over a saddle point, since $\pb \to {\bf 0}$ as the optimal trajectory approaches the saddle point \cite{Dykman1994d}.

In conclusion, we have considered the exponent in the extinction rate ${\cal Q}$ and demonstrated that it may be fragile. A small perturbation ($\propto \mu$) can change it significantly, ${\cal Q}$ for $\mu\to 0$ differs from ${\cal Q}$ for $\mu=0$. The fragility is related to the discontinuity of the quasi-stationary extinction current and the related steep slope of the quasistationary probability distribution near the extinction state. A formal condition for the onset of fragility is derived. Explicit results are obtained for the broadly used SIS model of epidemics, and it is shown that this model is fragile with respect to fluctuations of the total population. The analytical results are quantitatively confirmed by simulations.

We are grateful to A. Kamenev, I.B. Schwartz, and S. Shaw for stimulating discussions. The research was supported in part by the Army Research Office and by NSF grant No. PHY-0555346.

%\bibliography{C:/Aaa/BibTex/md10,C:/Aaa/Tdir/Largefl/Nongauss/Extinction/grand_bibtex,C:/Aaa/Tdir/Largefl/MishaKh}

 \end{document}